# ANALYSIS OF RECENT CHECKPOINTING TECHNIQUES FOR MOBILE COMPUTING SYSTEMS

Ruchi Tuli[1] and Parveen Kumar[2]

[1]Research Scholar, Singhania University (Pacheri Bari), Rajasthan (INDIA)
`tuli.ruchi@gmail.com`
[2]Professor, Meerut Institute of Engineering & Technology, Meerut (INDIA)
`pk223475@yahoo.com`

*Abstract*

*Recovery from transient failures is one of the prime issues in the context of distributed systems. These systems demand to have transparent yet efficient techniques to achieve the same. Checkpoint is defined as a designated place in a program where normal processing of a system is interrupted to preserve the status information. Checkpointing is a process of saving status information. Mobile computing systems often suffer from high failure rates that are transient and independent in nature. To add reliability and high availability to such distributed systems, checkpoint based rollback recovery is one of the widely used techniques for applications such as scientific computing, database, telecommunication applications and mission critical applications. This paper surveys the algorithms which have been reported in the literature for checkpointing in Mobile Computing Systems.*

*Keywords*– *Mobile computing systems, Co-ordinated checkpoint, rollback recovery, mobile host.*

## 1. Introduction

Checkpointing / rollback-recovery strategy has been an attractive approach for providing fault-tolerance to distributed applications. A checkpoint is a snapshot of the local state of a process, saved on local nonvolatile storage to survive process failures. A global checkpoint of an n-process distributed system consists of n checkpoints (local) such that each of these n checkpoints corresponds uniquely to one of the n processes. A global checkpoint M is defined as a consistent global checkpoint if no message is sent after a checkpoint of M and received before another checkpoint of M. The checkpoints belonging to a consistent global checkpoint are called globally consistent checkpoints (GCCs). In distributed systems, rollback recovery is complicated because messages induce inter-process dependencies during failure-free operation. Upon a failure of one or more processes in a system, these dependencies may force some of the processes that did not fail to roll back, creating what is commonly called rollback propagation. To see why rollback propagation occurs, consider the situation where the sender of a message m rolls back to a state that precedes the sending of m. The receiver of m must also roll back to a state that precedes m's receipt; otherwise, the states of the two processes would be inconsistent because they would show that message m was received without being sent, which is impossible in any correct failure-free execution. This phenomenon of cascaded rollback is called the *domino effect*. In some situations, rollback propagation may extend back to the initial state of the computation, losing all the work performed before the failure.

In a distributed system, if each participating process takes its checkpoints independently, then the system is susceptible to the domino effect. This approach is called i*ndependent or uncoordinated checkpointing* [1], [2], [3]. It is obviously desirable to avoid the domino effect and therefore several techniques have been developed to prevent it. One such technique is *coordinated checkpointing* [4], [5], [6] where processes coordinate their checkpoints to form a system-wide consistent state. In case of a process failure, the system state can be restored to such a consistent set of checkpoints, preventing the rollback propagation. Alternatively, *communication-induced checkpointing* [7], [8], [9] forces each process to take checkpoints based on information piggybacked on the application messages it receives from other processes. Checkpoints are taken such that a system-wide consistent state always exists on stable storage, thereby avoiding the domino effect. *Log-based rollback recovery* [10], [11], [12], [13], [14], [15] combines checkpointing with logging of nondeterministic events. Log-based rollback recovery relies on the piecewise deterministic (PWD) assumption, which postulates that all non-deterministic events that a process executes can be identified and that the information necessary to replay each event during recovery can be logged in the event's determinant. By logging and replaying the non-deterministic events in their exact original order, a process can deterministically recreate its pre-failure state even if this state has not been checkpointed. Table 1 below gives a comparison of rollback recovery protocols based on different parameters.

Table 1. - Comparison of Rollback recovery protocols

| Parameters | Uncoordinated Checkpointing | Coordinated Checkpointing | Communication Induced Checkpointing | Message logging protocols | | |
|---|---|---|---|---|---|---|
| | | | | Pessimistic logging | Optimstic logging | Casual logging |
| **Domino Effect** | Possible | No | No | No | No | No |
| **Orphan Message** | Possible | No | Possible | No | Possible | No |
| **Recovery Line** | Unbounded | Last global checkpoint | Possibly several checkpoints | Last checkpoint | Possibly several check points | Last check point |
| **Output Commit** | Not possible | Global Coordination required | Global Coordination required | Local decision | Global Coordination required | Local decision |

## 2. System Model

The algorithms that are considered in this paper use the common system model in which a mobile computing system consists of a set of mobile hosts (MHs) and mobile support stations (MSSs). The static MSS provides various services to support the MHs and a region covered by a MSS is called a cell. A wireless communication link is established between a MH and a MSS; and a high speed wired communication link is assumed between any two MSSs. The wireless links support FIFO communication in both directions between a MSS and the MHs in the cell. A distributed computation is performed by a set of MHs or MSSs in the network.

## 3. Checkpointing Algorithms for Mobile Computing Systems

Checkpointing techniques are studied under Asynchronous or uncoordinated, Synchronous or coordinated and quasi-synchronous or communication-induced checkpointing schemes. In this section, we discuss the various algorithms that have been proposed in literature for each of these schemes. Figure 1 shows the classification of these schemes.

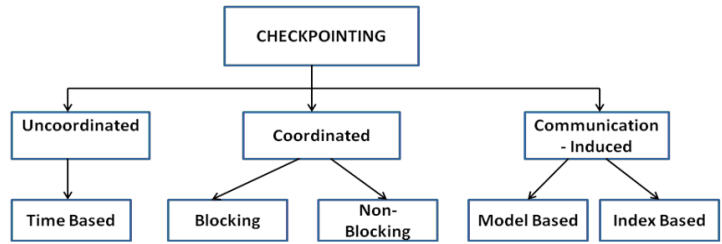

Fig. 1. – Classification of checkpointing schemes

### 3.1 Asynchronous or Uncoordinated or Independent checkpointing

In uncoordinated checkpointing, each process has autonomy in deciding when to take checkpoints. This eliminates the synchronization overhead as there is no need for coordination between processes and it allows processes to take checkpoints when it is most convenient or efficient.

Park, Woo and Ycom[16] proposed an algorithm based on independent checkpointing and asynchronous message logging. All the messages are delivered to mobile host (MH) through MSS, so message logs are saved by MSS for all MHs in its vicinity. The logs that are saved in MSS are used to recover state of process at MH after failure. Also, to reduce the message overhead, the mobile support stations take care of the dependency tracking.

Park, Woo and Ycom[17] proposed a scheme based on the message logging and independent checkpointing, and for the efficient management of the recovery information, such as checkpoints and message logs. They suggested a movement-based scheme which allows the movement of checkpoint and message logs to a nearby MSS when either distance between MH and MSS on which latest checkpoint is saved exceed a threshold value, or, when number of handoffs that number of MSS carrying message logs of a MH exceeds a threshold value. These schemes keep the recovery information of MH in certain range. The movement-based scheme considers both of the failure-free execution cost and the failure-recovery cost.

Zhang, Zuo, Zhi- Bowu and Yang [18] improved this scheme by migrating only partial recovery information of a MH when a MH moves out of the range. It means that recovery information of MH which is stored in some MSS due to mobility, is mapped to another set of MSSs. These MSSs are given by route function. The main advantage of this scheme is that one MSS is not burdened by transferring all the information to it.

Another movement-based algorithm was proposed by E. George, Chen and Jin [19] in which Independent checkpointing and optimistic message logging is used. MH takes checkpoint when its handoff_counter becomes greater than a predefined threshold.

### 3.2 Coordinated Checkpointing

In this we will discuss the algorithms for both blocking and non-blocking coordinated checkpointing schemes.

### 3.2.1 Blocking Coordinated Checkpointing

A straightforward approach to coordinated checkpointing is to block communications while the checkpointing protocol executes. After a process takes a local checkpoint, to prevent orphan messages, it remains blocked until the entire checkpointing activity is complete.

A two-level blocking checkpointing algorithm was proposed by Lotfi, Motamedi and Bandarabadi [20] in which local and global checkpoint are taken. Nodes take local checkpoint according to checkpoint interval calculated previously based on failure rate and save it in their local disk. These checkpoints when sent to stable storage become global checkpoint. Local checkpoints are used to recover from more probable failures where as global checkpoints are used to recover from less probable failures. After each checkpointing interval, system determines expected recovery time in case of permanent failure. System calculates amount of time taken (T1) to recover if system does not take global checkpoint and amount of time taken (T2) to recover if system takes global checkpoint. Then system compares these two times. If T2 < T1, system will take global checkpoint else system will only store checkpoint locally.

Awasthi and Kumar[21] proposed a synchronous checkpointing protocol for mobile distributed systems. They reduced the useless checkpoints and blocking of processes during checkpointing using a probabilistic approach. A process takes an induced checkpoint if the probability that it will get a checkpoint request in current initiation is high.

Another blocking coordinated scheme is proposed by Suparna Biswas and Sarmistha Neogy [22] in which each MSSp is required to maintain an array A[n] in which A[1] is 1 when MH1 is present in vicinity of cell of MSSp where number of MH (Mobile Host) are n starting from 0 to n-1. A MH initiates checkpointing procedure, calculates its dependency vector D and sends request to all the MH whose bit in dependency vector D is 1 via its MSS. If a MH is present in vicinity of current MSS, then checkpoint request is send directly to MH. Else current MSS will broadcast checkpoint request message to other MSS so that it can reach all those processes whose bit is 1 in dependency vector D calculated by checkpoint initiator. Thus all these processes take checkpoint and sends information to initiator via their local MSS.

Guohui Li and LihChyun Shu [23] designed an algorithm to reduce blocking time for checkpointing operation, in which each process Pi maintains a set of processes Si. A process Pj is included in this set if Pj has sent at least one message to Pi in current checkpoint interval. Checkpointing dependency information is transferred from sending process to destination process during normal message transmission. So when a process starts a checkpointing procedure, it knows in advance the processes on which it depends both transitively and directly.

Biswas & Neogy [24] proposed a checkpointing and failure recovery algorithm where mobile hosts save checkpoints based on mobility and movement patterns. Mobile hosts save checkpoints when number of hand-offs exceed a predefined handoff threshold value. They introduced the concept of migration checkpoint An MH upon saving migration checkpoint, sends it attached with migration message to its current MSS before disconnection.

S. Kumar, R.Garg [34] and P. Kumar [35] gave the concept of hybrid checkpointing algorithm, where in an all-process coordinated checkpoint is taken after the execution of minimum process coordinated checkpointing algorithm for a fixed number of times.

### 3.2.2. Non-Blocking Coordinated Checkpointing algorithm

In this approach the processes need not stop their execution while taking checkpoints. A fundamental problem in coordinated checkpointing is to prevent a process from receiving application messages that could make the checkpoint inconsistent.

Cao and Singhal [25] presented a non-blocking coordinated checkpointing algorithm with the concept of "Mutable Checkpoint" which is neither temporary nor permanent and can be converted to temporary checkpoint or discarded later and can be saved anywhere, e.g., the main memory or local disk of MHs. In this scheme MHs save a disconnection checkpoint before any type of disconnection .This checkpoint is converted to permanent checkpoint or discarded later. In this scheme only dependent processes are forced to take checkpoints. In this way, taking a mutable checkpoint avoids the overhead of transferring large amounts of data to the stable storage at MSSs over the wireless network.

Cao-Chen-Zhang-He [26] proposed an algorithm for Hybrid Systems. They presented an algorithm which was developed for integrating independent and coordinated checkpointing for application running on a hybrid distributed system containing multiple heterogeneous systems.

Bidyut – Rahimi- Liu [27] presented their work for mobile computing systems. In that work they presented a single phase non-blocking coordinated checkpointing suitable for mobile systems. This algorithm produces a consistent set of checkpoints without the overhead of temporary checkpoints.

Bidyut-Rahimi-Ziping Liu [28] proposed non-blocking checkpointing and recovery algorithms for bidirectional networks. The proposed algorithm allowed the process to take permanent checkpoints directly, without taking temporary checkpoint global snapshot algorithms for large scale distributed systems. Whenever a process is busy it takes a checkpoint after completing its current procedure. The algorithm was designed and simulate for Ring network.

Partha Sarathi Mandal and Krishnendu Mukhopadhyaya [29] proposed a non blocking algorithm that uses the concept of mobile agent to handle multiple initiations of checkpointing. Mobile Agent has id same as its initiator id and it migrates among processes, perform some work, take some actions and then moves to other node together with required information. Each process takes initial permanent checkpoint and sets version number of checkpoint to 0. Process sends application message m by piggybacking it with version number of its latest checkpoint. Receiver compares application message's version number with its own current checkpoint version number to decide whether to take checkpoint first or simply to process message only. There is a DFS which is maintained by each process which contains id of neighbors on which the process depends.

S. Kumar, R.K. Chauhan and P. Kumar [33] proposed a single-phase non-blocking coordinated checkpointing algorithm suitable for mobile computing environments in which processes take permanent checkpoints directly without taking temporary checkpoints and whenever a process is busy, the process takes a checkpoint after completing the current procedure.

### 3.3 Communication Induced or quasi-synchronous checkpointing

It lies between synchronous and asynchronous (independent) checkpointing. Process takes communication induced checkpoints besides independent checkpoint to reduce number of use-less checkpoints taken in independent checkpointing approach. Processes takes two kinds of checkpoints called local checkpoints and forced checkpoints. Local checkpoints are just like independent checkpoints taken in independent checkpointing approach. Forced checkpoints are taken to guarantee eventual progress of recovery line.

Qiangfeng Jiang and D. Manivannan [30] presented an optimistic checkpointing and selective message logging approach for consistent global checkpoint collection in distributed systems. In this work they presented a novel quasi-synchronous checkpointing algorithm that makes every checkpoint belong to a consistent global checkpoint. Under this algorithm every process takes tentative checkpoints and optimistically logs messages received after a tentative checkpoint is taken and before the tentative checkpoint is finalized. Since tentative checkpoint can be taken any time and sorted in local memory, tentative checkpoints taken can be flushed to stable storage anytime before that checkpoint is finalized.

Ajay D Kshemkalyani algorithm [31] presented a fast and message efficient algorithm and show that new algorithm is more efficient. He presented two new algorithms Simple Tree and Hypercube that use fewer message and have lower response time and parallel communication times. In addition the hypercube algorithm is symmetrical and has greater potential for balanced workload and congestion freedom. This algorithm have direct applicable in large scale distributed systems such as peer to peer and MIMD supercomputers

Jin Yang, Jiannong Cao, Weigang Wu [32] proposed a communication induced checkpointing scheme in which communication induced or forced checkpoints are taken by a process by analyzing piggybacked information that comes with received message. Each process has a logical clock or counter which is increased with every new checkpoint taken. When a process sends an application message, it piggybacks recent value of logical clock on message. Receiver compares its LC (logical clock) with received LC to decide whether to take a forced checkpoint before processing message or simply process the message. Algorithm uses a Mobile Agent (MA) system which has a globally unique id. Each MA executes on a node and takes an independent checkpoint before migration. It then determines next host to which it has to migrate, it reaches on that host and takes a checkpoint on it. This process will continue until all hosts have been visited. These checkpoints are called local checkpoints.

## 4. Application Scenarios

The checkpointing techniques discussed above have certain unique features which make them suitable to be used in a particular situation. Table 2 below discuses the application areas where each of these checkpointing techniques can be efficiently applied.

Table 2. : Application scenarios of checkpointing techniques

| Checkpointing Technique | Features | Application Area |
|---|---|---|
| Un-coordinated checkpointing | The Independent Checkpoint pattern is ideal for the development of systems with demanding performance constraints during error-free executions, that do not experience errors often and when they do they can afford to go off service for repairing the failure. | Telecom operator network and ISP service network |
| Co-ordinated | The Coordinated Checkpoint pattern is ad- | Automatic navigation |

| checkpointing | dressed more to the development of systemsthat have bounded time constraints (yet not high performance once) on their execution, and they cannot afford long execution delays due to system recovery. | control and embedded systems (e.g. mobile phones, PDAs) |
| Communication-induced checkpointing | The Communication-Induced Checkpoint pattern is meant for high-performance real time systems that can perform general purpose computations (as opposed to the systems that can perform only special purpose computations such as signal processing). | Stock market software |

## 5. Conclusion

We have reviewed and compared different approaches for failure free execution of a mobile host and to a greater extent failure free execution of mobile environment. We studied three checkpointing scheme- independent, coordinated and communication induced checkpointing and the various algorithms that have been developed under each of these scheme. Clearly, the higher the level of abstraction provided by a communication model, the simpler the snapshot algorithm. The requirement of global snapshots finds a large number of applications like: detection of stable properties, checkpointing, monitoring, debugging, analyses of distributed computation, discarding of obsolete information, etc. We have also shown the features that are needed to be considered while choosing a checkpointing technique for a particular system.